\documentclass[twocolumn,english,aps,prl,showpacs,superscriptaddress,
preprintnumbers]{revtex4}
\usepackage[T1]{fontenc}
\usepackage[latin9]{inputenc}
\usepackage{amsmath}
\usepackage{amssymb}
\usepackage{graphicx}

\makeatletter

\usepackage{amstext}

\usepackage{babel}

\makeatother

\begin{document}
\title{Tunable Graphene Antennas for Selective Enhancement of THz-Emission}
\author{R. Filter}
\homepage{http://www.robertfilter.de}
\affiliation{
Institute of Condensed Matter Theory and Solid State Optics,
Abbe Center of Photonics, Friedrich-Schiller-Universität Jena, D-07743
Jena, Germany}
\author{M. Farhat}
\affiliation{
Institute of Condensed Matter Theory and Solid State Optics,
Abbe Center of Photonics, Friedrich-Schiller-Universität Jena, D-07743
Jena, Germany}
\affiliation{Division of Computer, Electrical, and Mathematical Sciences and Engineering,
King Abdullah University of Science and Technology (KAUST), Thuwal, 23955-6900, Saudi Arabia}
\author{M. Steglich}
\affiliation{Laboratory Astrophysics Group of the Max Planck Institute for
Astronomy, Friedrich-Schiller-Universität Jena, D-07743 Jena, Germany}
\affiliation{Department of Chemistry, University of Basel, Klingelbergstrasse 80, CH-4056 Basel, Switzerland}
\author{R. Alaee}
\affiliation{
Institute of Condensed Matter Theory and Solid State Optics,
Abbe Center of Photonics, Friedrich-Schiller-Universität Jena, D-07743
Jena, Germany}
\author{C. Rockstuhl}
\affiliation{
Institute of Condensed Matter Theory and Solid State Optics,
Abbe Center of Photonics, Friedrich-Schiller-Universität Jena, D-07743
Jena, Germany}
\author{F. Lederer}
\affiliation{
Institute of Condensed Matter Theory and Solid State Optics,
Abbe Center of Photonics, Friedrich-Schiller-Universität Jena, D-07743
Jena, Germany}

\begin{abstract}
In this paper, we will introduce THz graphene antennas that
strongly enhance the emission rate of quantum systems at specific frequencies.
The tunability of these antennas can be used to
selectively enhance individual spectral features.
We will show as an example that any weak transition in the spectrum of
coronene can become the dominant contribution.
This selective and tunable enhancement
establishes a new class of graphene-based THz devices, which will find
applications in sensors, novel light sources, spectroscopy, and quantum
communication devices.
\end{abstract}

%\ocis{(020.4900) Oscillator strengths; (300.2140) Emission; (300.6390) molecular Spectroscopy; (300.6495) terahertz Spectroscopy.}
\maketitle

\section{Introduction}

The advance of optical antennas has led to incredible new
possibilities to control light-matter-interactions; including the directive
emission of quantum systems or the modification of radiative rates at
which such hybrid systems emit light.\cite{Muhlschlegel2005,vanHulst2010,Anger2006}
The parameter usually used to quantify modified emission rates of quantum systems
is the Purcell factor $F$ defined as
$F=\gamma_{\mathrm{rad}}^{\mathrm{a}}/\gamma_{\mathrm{rad}}^{\mathrm{fs}}$
with the radiative emission rate $\gamma_{\mathrm{rad}}$. The superscripts
``a'' and ``fs'' denote the presence of the antenna or an emission in free space,
respectively. For a
cavity, $F$ is proportional to the quality factor $Q$ and $\lambda^{3}/V$,
where $\lambda$ is the wavelength and $V$ is the mode volume. High
Purcell factors have been achieved either in high-Q resonators, tolerating larger
mode volumes,\cite{Song2005} or in systems supporting localized surface
plasmon polaritons (LSPP) with extremely small mode volumes, therefore
tolerating lower Q-factors \cite{Campillo1991,Greffet2010,Koppens2011,Filter2011}.
However, since comparable Purcell factors appear for modes at
higher frequencies in most of such implementations, these approaches
are inappropriate for the selective enhancement of emissions at individual
frequencies exclusively.
Moreover, the impossibility to tune the spectral
response of metal-based antennas over extended spectral domains is
extremely restrictive for many applications.
Mechanically tunable metallic cavities are however able to
selectively enhance electric \textit{or} magnetic dipole transitions \cite{Zia2011}.

\begin{figure}
\begin{centering}
\includegraphics[keepaspectratio=true,width=8cm]{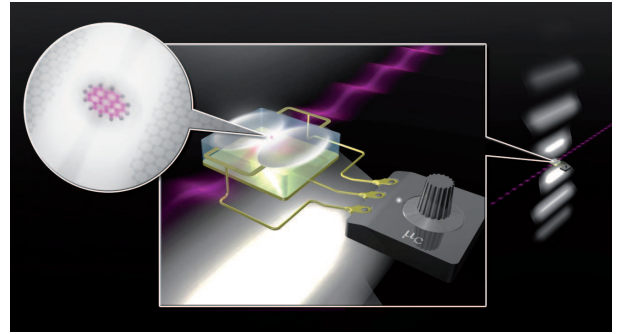}
\par\end{centering}
\caption{\label{fig:fancy}
A graphene antenna to selectively enhance THz emissions: A molecule gets excited in the visible/UV.
Subsequentially, it strongly emits at a single frequency in the THz regime. The emission frequency
can be tuned by adjusting the chemical potential $\mu_c$.}
\end{figure}

Recently, graphene has been suggested as a possible material that may
remedy at least some of these limitations \cite{Koppens2011}.
Moreover, since its operational domain would be in the THz frequency range
($\nu\approx0.1-30$~THz), it may also be used for long-needed functional devices
in the so-called THz gap \cite{Tamagnone2012a,Tamagnone2012b,Rasoul2012}. An example is the
graphene antenna discussed in this work and visualized in Fig.~\ref{fig:fancy}.
In the THz frequency domain,
graphene offers low-loss SPPs \cite{Boehm1962,Novoselov2004,Neto2009,fei2011,chen2012},
which have a lower loss than their metallic counterparts for
two reasons, although graphene's advantage at this point is not extraordinary.
First, the intrinsic losses are less since the electronic scattering rate is roughly one order of magnitude lower than Drude loss rates. Second, the one-atom thick structure has a smaller cross section to the actual SPP mode.
The effective wavelengths of propagating SPPs supported by graphene
are two to three orders of magnitude smaller than the free space wavelengths
at the same frequency \cite{Hanson2008}.
The small wavelengths of the propagating SPP's on graphene are in
turn implying small resonant structures supporting localized SPP's which
are needed for high Purcell factors \cite{hasan2011}.
Most notably, the electronic properties of graphene can
be widely tuned using the electric field effect \cite{Yu2009,Ju2011,yan2012}.
A final advantage of graphene is that it does not support SPPs at optical frequencies \cite{jablan2009}. This property is extremely useful for the selective enhancement of THz emission bands since optical emission
bands are only marginally influenced by graphene antennas, i.e. one may assume $F_{\mathrm{rad}}\left(\omega_{\mathrm{optical}}\right)\approx 1$.

A referential example for an application in the THz domain is the
identification of specific molecules, e.g., in any transmission, fluorescence,
or Raman spectroscopy experiments which rely on previously collected
spectroscopic information \cite{Schumacher2011}. Although
graphene-based optical antennas were already suggested to support such
processes\cite{Koppens2011,Xu2012}, in principle, metal-based antennas
can also be used in such applications. Therefore, the two key
questions to be answered are:
\textit{How can graphene
antennas be used to observe effects which are fundamentally inaccessible by metal
antennas and how can basic limitations of the latter ones be overcome?}

Here we introduce a concept that allows the selective enhancement
emissions in molecules using tunable THz antennas made of graphene. It
will be shown that distinct and very weak spectral features at disparate
frequencies can be enhanced without any structural modification of the
antenna. The scheme we are going to introduce provides the possibility to
probe for all spectral features of a certain molecule; thus
simplifying the spectroscopic classification of the molecule tremendously. To enhance
spectrally weak features, it is essential that the enhancement is selective,
i.e. no other transitions shall be strongly enhanced except the desired one.
This requires carefully designed antennas where higher-order resonances
are shifted off the spectral domain of interest.

For this reason, we will discuss the optical properties of a
graphene-based antenna and elaborate on the details of the ultimate design first. All spectral
properties are clearly motivated by the requirements described above.
We will also discuss the interaction of the antenna with a specific molecule
and outline how its spectral properties can be selectively enhanced.

\section{Efficient and tunable graphene antennas for strong spontaneous emission enhancement}

As mentioned earlier, graphene SPPs are prone to losses.
Therefore, the rate at which radiation can escape an antenna-emitter
system $\gamma_{\mathrm{rad}}$ is lower than
the total emission rate $\gamma_{\mathrm{tot}}$ of the emitter,
$\gamma_{\mathrm{tot}}=\gamma_{\mathrm{loss}}+\gamma_{\mathrm{rad}}$.
Both quantities are related by the antenna efficiency
$\eta=\gamma_{\mathrm{rad}}/\gamma_{\mathrm{tot}}$.
The enhancement $F_\mathrm{tot}$ of the total emission rate $\gamma_{\mathrm{tot}}$ of a dipole emitter
in the vicinity of an antenna compared to the case of free space
can be calculated within the Weisskopf-Wigner approximation as
$F_{\mathrm{tot}} = \gamma_{\mathrm{tot}}^{\mathrm{a}}/\gamma_{\mathrm{tot}}^{\mathrm{fs}}=
P_{\mathrm{tot}}^{\mathrm{a}}/P_{\mathrm{tot}}^{\mathrm{fs}} $
\cite{Welsch2006}.
%\begin{eqnarray}
%F_{\mathrm{tot}} & = & \frac{\gamma_{\mathrm{tot}}^{\mathrm{a}}}{\gamma_{\mathrm{tot}}^{\mathrm{fs}}}=\frac{P_{\mathrm{tot}}^{\mathrm{a}}}{P_{\mathrm{tot}}^{\mathrm{fs}}}\ .\label{eq:total_rate_enhancement-1}
%\end{eqnarray}
Here, $P_{\mathrm{tot}}^{\mathrm{a/fs}}$ is the total power emitted by
the dipole. Using the latter relation, it is possible to calculate the
emission rates in the realm of classical electrodynamics. Assuming that the
emission in free space is free of absorption ($\eta^{\mathrm{fs}}=1$), the
Purcell factor with respect to the antenna reads as
$F=\eta^{\mathrm{a}}F_{\mathrm{tot}}$. In general, $F_{\mathrm{tot}}$ depends
on the position of the dipole, its polarization, and all properties of the
antenna.

In the following, we will show how to
achieve a high Purcell factor for a graphene-based antenna in the THz
regime at a preselected frequency exclusively.
The electric field effect can be used to change Graphene's chemical potential $\mu_c$
and consequentially its conductivity. See App.~\ref{sec:graphene-numerics} for
more information and details on our numerical implementation of the material.
We further assume that the devices required to apply the electric field effect
do not strongly influence the overall emission properties.
This simplifies the analysis to free standing elements.

The design of plasmonic antennas with
high efficiency $\eta^a$ at a certain frequency strongly depends on
geometrical parameters. Notably, the efficiency of
sub-wavelength antennas can be quite low.
To achieve a high Purcell factor, one thus encounters a trade-off between
high $F_\mathrm{tot}$ and $\eta^a$.
Plasmonic antennas consisting of two elements generally exhibit
a larger efficiency than single element antennas \cite{Rogobete2007}.
In consequence, a configuration is considered where the dipole is
situated in the center between the two elements of the antenna.
The dipole is oriented in $x$-direction, i.e. parallel to the connection line of both elements.
\begin{figure}
\begin{centering}
\includegraphics[width=8cm]{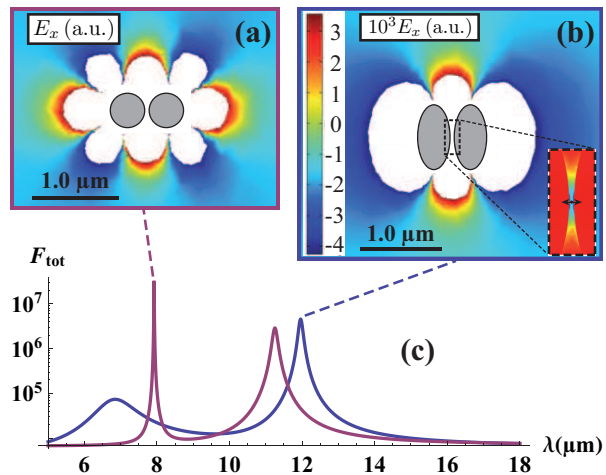}
\par\end{centering}
\caption{\label{fig:Field Plot-1}
(a) \& (c) Two circular graphene elements cause a strong enhancement of the total emission rate
$F_\mathrm{tot}$ for
a dark antenna mode at $\lambda\approx 7.9$~$\mu$m ($\nu\approx38$~THz). To visualize the field
outside the antenna, the colormap has been truncated (white regions).
(b) \& (c) The antenna with two elliptical elements exhibits by far the strongest
$F_\mathrm{tot}$ for the desired dipolar
mode at $\lambda\approx 12$~$\mu$m ($\nu\approx26.6$~THz).
The scale shown in (b) applies for all field plots with order-of magnitude rescalings.}
\end{figure}
As a result of the strong confinement of the LSPP,
a small separation between the antenna elements is required.
Initially we chose an antenna design consisting of two circular elements.
Such a device can be fabricated e.g. using nanosphere lithography and the
single discs may be described as Fabry-Perot resonators \cite{Cong2009,FilterCircular}.
A configuration with radii of $200\,$nm
and separation of $30\,$nm was found to exhibit
$F_{\mathrm{tot}}\approx10^{6\dots7}$, and an efficiency
$\eta\approx0.32 $ at $\nu\approx26.6\,$THz with
$\mu_{c}=1\,$eV. However, this structure also permits the coupling of the
dipole to higher-order resonances of the antenna, see Fig.~\ref{fig:Field
Plot-1}(a). These higher-order resonances, sometimes called dark modes,
have extremely low efficiencies since they are only weakly coupled to the
far-field. Nevertheless, they exhibit a strong $F_\mathrm{tot}$ and are
therefore undesired. For our
present scenario, a dark mode at $\nu\approx38\,$THz had an
enhancement exceeding that of the dipole resonance with an efficiency of
only $\eta\approx0.035$. Therefore, these higher-order resonances have to be
suppressed in order to render the device useful.

We found that the problem can be
circumvented by stretching the discs in the vertical ($y$) direction, i.e.
perpendicular to the dipole, by a factor of two. This stretching
results in an increase of the effective radiating area of the antenna and
thus of its efficiency. This is accompanied by a blueshift of the undesired
higher-order modes \cite{Rogobete2007}. We observed a suppression of
these dark modes by several orders of magnitude for an antenna made of
two elliptical elements. Another aspect in favor of our design is a
further enhancement of $F_{\mathrm{tot}}$, see
Fig.~\ref{fig:Field Plot-1}(c). Generally,
$F_{\mathrm{tot}}$ at other frequencies than the antenna's dipole
resonance is two or more orders of magnitude smaller. This offset
permits the desired selective enhancement of molecular resonances if the
resonance of the antenna coincides with one of the resonances of a
quantum system as it will be outlined in the following section.

\begin{figure}
\begin{centering}
\includegraphics[width=8cm]{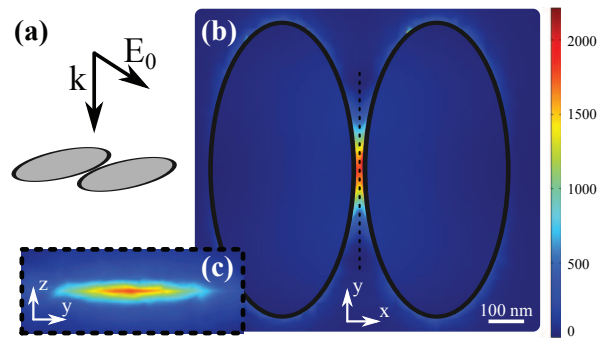}
\par\end{centering}
\caption{\label{fig:modelocalization}
(a) A plane wave with amplitude $E_0$ excites the dipolar resonance $\lambda\approx 12\,\mu$m of the antenna.
(b) \& (c): Distribution of $\left| \mathbf{E}\left(\mathbf{r}\right)/E_0 \right| $ in the plane of the graphene ellipses (x-y-plane) and in-between at $x=0$ (y-z-plane).
The region with strong enhancement is approximately $30\times100\times20\,\text{nm}^3$ (blue region).
For a dipole situated here, the Purcell effect is comparable to a placement in the center.}
\end{figure}

One may furthermore ask the question how strongly localized the Purcell enhancement
is for the found efficient antenna. Because of the plasmonic nature of the mode, one
expects a rather small region of enhancement. To measure the size of this region,
one can use the reciprocity theorem:\cite{Balanis2005,Esteban2009,Bharadwaj2009}
Illuminating the antenna with a plane
wave corresponds to an emitting dipole at infinity.
We assume that the plane wave excitation
comes from a direction of strong emission of the antenna's dipole mode.
Then, the field enhancement at some position
is a measure for the radiated power of a dipole at that very position.
We further
assume that the dipole emitter couple to the dipole mode of the antenna.
Then, radiation patterns do not change strongly for different emitter positions.
Consequentially, the ratios of field enhancements are proportional to the ratios of the overall
radiated power and thus the ratio of Purcell enhancements. We have
made a corresponding calculation. As expected, we found
a region of strong interaction confined in-between the antenna, see Fig.~\ref{fig:modelocalization}.

\section{The changed spectrum of coronene}
Up to this point, the enhancement of the total emission rate $F_{\mathrm{tot}}$
was referring to a dipole in the vicinity of a
graphene antenna. To demonstrate the performance
of the suggested implementation, coronene ($\text{C}_{24}\text{H}_{12}$)
has been chosen as an example.
It belongs to the group of polycyclic aromatic
hydrocarbon (PAH) complexes and is of both great practical and theoretical
interest \cite{Tielens2005}. The molecule exhibits resonances in the THz
regime and its properties are well-documented \cite{Langhoff1996,Steglich2010}.

The emission process of coronene can be understood as a
stepwise procedure following Ref.~\cite{Tielens2005}. First,
upon absorption of a UV photon, the electronic configuration of the
molecule is transferred into an excited state.
The excitation will be assumed in the
near ultraviolet around a frequency of $\sim990\,$THz ($300\,$nm)
where coronene absorbs strongly. The absorbed energy is then almost
instantaneously redistributed to the different vibrational modes of the
molecule. The characteristic time scale of this process is about
$10^{-12}\,$s. This redistribution obeys a Planck distribution
$B\left(\omega;T_{\mathrm{m}}\right)$ over the respective eigenenergies
$\hbar\omega_{i}$. The temperature $T_{\mathrm{m}}$ refers to the
microcanonical temperature of the molecule and can be approximated by
$T_{\mathrm{m}}\simeq2000\left(E\left(eV\right)/N_{\mathrm{c}}\right)^{0.4}K$
with the energy of the exciting radiation $E$ in electronvolts and
the number $N_{\mathrm{c}}=24$ of carbon atoms of coronene.
The expression for $T_{\mathrm{m}}$ can be derived from the
density of states of PAH's. It
has been proven correct within an error of $\approx10\%$ for
$35\dots1000\,$K \cite{Tielens2005}. For coronene and the given
excitation, we get $T_{\mathrm{m}}\approx990\,$K. The emission
spectrum in free space is then given by the oscillator strengths
$f\left(\omega_{i}\right)$ weighted by $B\left(\omega;T_{m}\right)$
 \cite{Tielens2005}. This behaviour corresponds to the
generalized Planck radiation law for luminescence and holds for
a large class of molecules, semiconductors etc.; see e.g. Ref. \cite{wurfel1995}.

Close to the antenna, the emission of the molecule is strongly modified. We assume here,
that the redistribution of energy into the different eigenmodes is
much faster than any involved radiation process.
This seems reasonable since the natural emission time for THz radiation is in the order
of or even bigger than one second \cite{Tielens2005}.
Therefore, one might
hypothesize that even an enhancement of several orders of magnitude for
the given THz transitions in coronene should be possible without any
noticeable influence on the internal redistribution dynamics.
Hence, by enhancing
the emission rate for just one eigenmode, the emission of the
remaining ones decreases. Then, one finds different
total emission rate enhancements for each transition
$\mathcal{F}_{\mathrm{tot}}\left(\omega_{i}\right)$, which deviate from the
Purcell factors of a single dipole, i.e.
$\mathcal{F}_{\mathrm{tot}}\left(\omega_{i}\right)\neq
F_{\mathrm{tot}}\left(\omega_{i}\right)$. For a linear multiresonant
system, modified emission rates
are related to each other by a sum rule \cite{Barnett1996}.
Because of the internal redistribution dynamics of coronene,
this approach cannot be used.
We assume that the power emitted upon excitation does not
change if the antenna is present, i.e.,
$P_{\mathrm{tot}}^{\mathrm{a}}=
\sum_{i}\hbar\omega_{i}\gamma_{i}\mathcal{F}\left(\omega_{i}\right)
B\left(\omega_{k};T_{\mathrm{m}}\right)\overset{!}{=}P_{\mathrm{tot}}^{\mathrm{fs}}$
implying energy conservation. Then, the enhancement of the radiative rate with
respect to the antenna efficiency is given by
\begin{eqnarray}
\mathcal{F}_{\mathrm{rad}}\left(\omega_{i}\right) & = & \eta\left(\omega_{i}\right)
\cdot F_{\mathrm{tot}}\left(\omega_{i}\right)\cdot\nonumber \\
 &  & \frac{\sum_{k}\hbar\omega_{k}\gamma_{\mathrm{tot}}^{\mathrm{fs}}\left
 (\omega_{k}\right)B\left(\omega_{k};T_{\mathrm{m}}\right)}{\sum_{k}\hbar\omega_{k}
 F_{\mathrm{tot}}\left(\omega_{k}\right)\gamma_{\mathrm{tot}}^{\mathrm{fs}}
 \left(\omega_{k}\right)B\left(\omega_{k};T_{\mathrm{m}}\right)}\ .\ \label{eq:Molecular_Changed_Rates-1}
\end{eqnarray}
Since the characteristics of the graphene antenna can be tuned by
changing the chemical potential $\mu_{\mathrm{c}}$, the enhancement
also depends on $\mu_{\mathrm{c}}$. If now a certain transition is
selectively enhanced to a large extent,
$F_{\mathrm{tot}}\left(\omega_{i}\right)\gg
F_{\mathrm{tot}}\left(\omega_{k}\right)\ \left(k\neq i\right)$, the energy
supplied to the molecule will predominantly leave it by exactly this
transition. This result would also hold with respect to the sum rule in
Ref.~\cite{Barnett1996}, for which we would find
$\mathcal{F}_{\mathrm{tot}}^{\mathrm{BSL}}\left(\omega_{i}\right)=
F_{\mathrm{tot}}\left(\omega_{i}\right)/\sum_{k}F_{\mathrm{tot}}\left(\omega_{k}\right)$.
Hence, it might be expected that the shift of emission processes in favor of a single
transition by a selective enhancement is likely to be a generic property of molecular systems.

A temperature $T$ of the environment also changes the emission rates of molecules \cite{joulain2005,jones2012}. Then the condition for selective enhancement changes to $F_{\mathrm{tot}}\left(\omega_{i}\right)\cdot B\left(\omega_{i};T\right)\gg F_{\mathrm{tot}}\left(\omega_{k}\right)B\left(\omega_{k};T\right)\ \left( k\neq i\right) $. Hence, the selective enhancement is working for all frequencies $\omega_i$ if the difference in $F_{\mathrm{tot}}$ is much more significant than the differences of temperature induced emissions, which is generally the case for a coupling to resonant systems \cite{joulain2005}.

The spectral resolution of the selective enhancement is limited by the antenna.
It is naturally linked to the width of the antenna resonance which is approximately the relaxation rate $\Gamma$. So, if $\Gamma$ is smaller
than the minimum spectral distance between certain resonances one wishes to distinguish,
the selective enhancement should work. This is the case for coronene;
current fabrication techniques allow relaxation rates in the order of a few meV or THz, respectively \cite{Ju2011}.

\begin{figure}
\begin{centering}
\includegraphics[width=8cm]{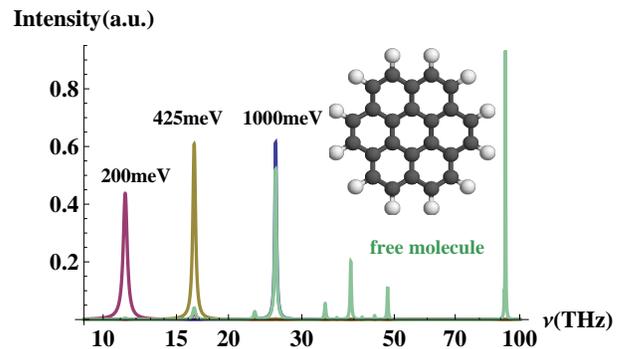}
\par\end{centering}
\caption{\label{fig:Changed_Spectrum_Coronene-1} The modified emission spectrum
of coronene in the vicinity of the graphene antenna as a function of the chemical potential $\mu_c$. For reference, the green curve
displays the emission spectrum of the bare molecule exhibiting several
peaks with different strengths.
When tuning the chemical potential by taking advantage of the electric field effect, the antenna may
strongly \textit{and} selectively
enhance only a single emission line, so that most of the energy leaves the
molecule via this transition.}
\end{figure}

To demonstrate this selective \textit{and} tunable enhancement,
Fig.~\ref{fig:Changed_Spectrum_Coronene-1} shows how the emission
spectrum of coronene changes if the antenna resonances are tuned. Although extensive
simulations were performed, only a few selective results are shown where
the resonance was tuned to enhance the emission at frequencies
corresponding to "jumping-jack"-, "drumhead"- and a
"C-H-out-of-plane"-vibrational motion at $11.3$~THz, $16.5$~THz and
$26.0$~THz, respectively \cite{Mattioda2009}. Noteworthy, these
emissions are electric dipole-allowed, consistent with our analysis of
the graphene antenna.
The quantitative
visualization of the spectra is performed by assuming, as usual, a certain
line width for each resonance. For PAH's, we used a linewidth of
$0.15\,$THz, which is characteristic for molecules isolated from each other \cite{Mattioda2009}.
It can be clearly seen that
by tuning the antenna it is possible to choose the transition which is
forced to emit light into the far-field; the main goal of this contribution.
Various effects might
broaden the lines as e.g. the coupling to surfaces or collisions with other
molecules. Nevertheless, it can be safely anticipated that the general
characteristics of the spectrum survive.

\section{Conclusion}
In conclusion, we have suggested and verified that due to the unique
properties of graphene, antennas made from this material can be used to
selectively enhance different molecular transitions at THz frequencies by
several orders of magnitude. The achievable efficiencies and the
suppression of undesired modes were found to strongly depend on the
actual antenna geometry. A suitable implementation which consists of two
closely placed elliptical graphene elements was introduced. Using this antenna,
weak transitions can be made dominant in the emission
spectrum. The potential of graphene antennas to selectively enhance
molecular transitions can clearly be anticipated to lead to new applications
in THz spectroscopy but will also contribute to the development of novel
sensors, highly directed single photon light sources, and quantum
communication and quantum computation devices \cite{Rockstuhl2009}.

\begin{appendix}
\section{Numerical implementation of graphene}\label{sec:graphene-numerics}

At temperature $T$, the in-plane conductivity of graphene can be
approximated using the Kubo formula
\begin{eqnarray}
\sigma_s\left(\omega\right) & = & \mathrm{i}\frac{1}{\pi\hbar^{2}}\frac{e^{2}k_{B}T}{\omega+\mathrm{i}2\Gamma}\left\{ \frac{\mu_{c}}{k_{B}T}+2\ln\left[\exp\left(-\frac{\mu_{c}}{k_{B}T}\right)+1\right]\right\} \nonumber \\ &  & +\mathrm{i}\frac{e^{2}}{4\pi\hbar}\ln\left[\frac{2\left|\mu_{c}\right|-\hbar\left(\omega+\mathrm{i}2\Gamma\right)}{2\left|\mu_{c}\right|+\hbar\left(\omega+\mathrm{i}2\Gamma\right)}\right]\label{eq:Graphene_conductivity-1}
\end{eqnarray}
with the chemical potential $\mu_{c}$ and the relaxation rate $\Gamma$ \cite{Hanson2008,Koppens2011}.
$\Gamma$ was taken to be $0.1\,$meV, consistent
with values reported earlier \cite{Vakil2011}.
$\Gamma$ may also be calculated from experimental data for the transport mobility \cite{Mayorov2011}.
Note that the second interband
term is not temperature dependent for $k_B T \ll \left|\mu_c\right|$.
This is in very good approximation for the values of the chemical potential
used in our analysis, even at room temperature, where $k_B T \approx 26\,$meV.
Throughout, we assumed a time dependency according to $\exp\left(-\mathrm{i}\omega t\right)$.
In our calculations, we represented graphene as a thin
conductive layer with thickness $d=1\,$nm $\ll\lambda$ and relative
permittivity $\varepsilon_{g}\left(\omega\right) =
1+\mathrm{i}\sigma_s\left(\omega\right)/ \left(\varepsilon_{0}\,\omega\,
d\right)$, see also
Refs.~\cite{Slepyan1999,Gusynin2007,Hanson2008,Chen2011,Koppens2011}
and \cite{Vakil2011}. It was verified that none of the results are
quantitatively affected by that finite thickness. All simulations were
performed using the Finite-Element Method as implemented in
COMSOL Multiphysics 3.5.
It may be explicitly noted that we take any non-radiative loss into account via
the losses in the graphene antenna and through the internal redistributions in coronene.
\end{appendix}

\section*{Acknowledgements}
We thank Karsten Verch (www.karstenverch.com) for his artistic view of
the theory in Fig.~\ref{fig:fancy}. We also appreciate the professional feedback and further
input of the anonymous reviewers to state certain points more clearly.
Financial support by the German Federal Ministry of Education and
Research (PhoNa), by the Thuringian State Government (MeMa) and the
German Science Foundation (SPP 1391 Ultrafast Nano-optics) is
acknowledged.
\bibliographystyle{apsrev4-1}
\bibliography{THzGrapheneAntenna_Bib_final}
\end{document}